\documentclass[12pt]{article}
\usepackage{amsmath,amssymb}
\usepackage{graphicx,color}
\numberwithin{equation}{section}
\usepackage{cite}
\usepackage{bm}
\usepackage{dcolumn}  
\newcommand{\be}{\begin{equation}}
\newcommand{\bea}{\begin{eqnarray}}
\newcommand{\eea}{\end{eqnarray}}
\newcommand{\ba}{\begin{array}}
\newcommand{\ea}{\end{array}}
\newcommand{\ee}{\end{equation}}

\expandafter\ifx\csname mathbbm\endcsname\relax

\else

\fi \textheight 22cm \textwidth 15cm \topmargin 1mm
\def\a{\alpha}
\def\a{\alpha'}
\begin{document}
\begin{titlepage}
\hfill
\vbox{
    \halign{#\hfil         \cr
           hep-th/0604106  \cr
           IPM/P-2006/022 \cr
           } 
      }  
\vspace*{20mm}
\begin{center}
{\Large {\bf $R^4$ Corrections to D1D5p Black Hole Entropy from Entropy Function Formalism}\\ }

\vspace*{15mm} 
\vspace*{1mm} 
{\large Ahmad Ghodsi}
\vspace*{1cm}

{\it Institute for Studies in Theoretical Physics and Mathematics (IPM)\\
P.O. Box 19395-5531, Tehran, Iran}\\
{\it and}\\
{\it {Centre de Physique Th\'eorique \'Ecole Polytechnique\\
91128 Palaiseau, France} }\\
\vskip 0.6 cm

E-mail: {{\it ahmad@ipm.ir}} \\
\vspace*{1cm}
\end{center}

\begin{abstract}
We show that in IIB string theory and for $D1D5p$ black holes in ten dimensions the method of entropy function works. Despite the more complicated Wald formula for the entropy of $D1D5p$ black holes in ten dimensions, their entropy is given by entropy function at its extremum point. We use this method for computing the entropy of the system both at the level of supergravity and for its higher order $\a^3R^4$ corrections.  
\end{abstract}

\end{titlepage}

\section{Introduction}

Black holes in string theory can be constructed out of various $D$-brane configurations. This is a well known result that the entropy of these black holes can be computed from the Wald formula \cite{Wald}, \cite{Visser:1993qa}. Further more it has been shown that this entropy is in exact agreement with counting the degeneracy of the microstates of such a configurations \cite{Strominger:1996sh},\cite{Cvetic}. For example recently new developments in the context of electrically charged heterotic black holes in four dimensions has been occurred (see \cite{Dabholkar}). 

Sen has been shown that that for a specific class of black holes, entropy is given by an entropy function at its extremum point \cite{Sen:2005wa}. This method can be summarized into a few steps:
 
1. Consider an extremal charged black hole with near horizon geometry $AdS_2\times S^{d-2}$ with constant radii $v_i$. The assumption is that this background is a solution to the equations of motion for a specific action. In this action gravity is coupled to a set of electric and magnetic fields and some neutral scalar fields $u_s$. 

2. Introducing a function which is the integral of Lagrangian density over the horizon $S^{d-2}$. This is a function of $v_i$ and $u_s$, the scalars of the theory, and electric and magnetic field strengths $(e_i, p_i)$.

3. Perform the Legendre transformation of this function with respect to electric field strengths $e_i$ and extremize it with respect to the scalars $v_i$ and $u_s$. This function (entropy function) at its extremum point is proportional to the entropy of black hole. 

In fact, this is a useful method for computing the entropy from the Wald formula. In addition, one can easily find the corrections to the entropy function. These properties suggest that one try to develop this method for other backgrounds and specially find a generalization for entropy function. The very simple nature of computations makes it possible to study easily the higher derivative corrections to entropy and also corrections to background at near horizon (see \cite{Sen:2005iz} for more recent papers). In this paper we decided to show that in the context of IIB string theory in ten dimensions and for special case of $D1D5p$ black holes this method works. 

The near horizon geometry of $D1D5p$ is $M_3\times S^3\times T^4$ where $M_3$ is deformed (boosted) $AdS_3$ geometry. In this system $D1$-branes play the role of sources for electric fields and $D5$-branes behave as sources for magnetic fields. Because of special form of this geometry and specially the existence of an off diagonal term in the metric finding the entropy from Wald formula is somehow seems complicated specially when higher derivative corrections take into the account. But it is possible to show that by reducing IIB action from ten dimensions over $S^1\times T^4$ to five dimensions the background is simplifying to $AdS_2\times S^3$ and we can use the entropy function method for it, although finding the reduced Lagrangian for higher order correction may not be so simple. 

Despite this fact, we are interested to know how this mechanism works in more complicated backgrounds in ten dimension. In addition we are interested to know the higher derivative corrections to entropy of these black holes. These are $\a$ corrections for low energy effective action of IIB theory where come from string three level scattering amplitude computations. In the case of IIB theory these corrections are of the order $\a^3R^4$ \cite{Grisaru:1986vi}. Although there are some ambiguities in the effective action due to field redefinitions but we will show that, the final result is independent of these ambiguities.  

In what follows we will show that despite the complicated form of Wald formula, the entropy of this system in a nontrivial way is given by entropy function, which is the Legendre transform with respect to electric field strength, from the integral of IIB action over the horizon. This suggests that the property of existence of entropy function is inherited when we reduce the background specially when the reduced geometry is $AdS_2\times S^{d-2}$.

The organization of this paper is as follows. In section two we briefly introduce $D1D5p$ black holes as a solution for IIB supergravity equations of motion. In section three starting from the reduced action $IIB/S^1\times T^4$ we review the method of entropy function for $D1D5p$ black holes in five dimensions. In section four we present a systematic way similar to the Sen's method for finding the entropy function for this special case of black holes in ten dimensions. Finally we compute $\a$ corrections to the entropy for $D1D5p$ by using the corrected entropy function. In last section we discuss about the method the results.

\section{$D1D5p$ black holes in IIB supergravity}
Before starting to compute the entropy function for $D1D5p$ black hole let's review this system as a solution to the equations of motion for IIB supergravity. We Begin from IIB supergravity action in string frame
\be
S_{IIB}=\frac{1}{16\pi G_{D}}\int d^{D}x\sqrt{-g}\bigg\{ e^{-2\phi}(R+4\partial_\mu\phi\partial^\mu\phi)-\frac12\sum_n\frac{1}{n!}F_n^2\bigg\}\,,\label{IIB}
\ee
where we have just turned on the RR charges. The equations of motion are
\bea
R_{\mu\nu}+4\partial_\mu\phi\partial_\nu\phi&=&e^{2\phi}\sum_n\frac{1}{2n!}(nF_{\mu\alpha_2...\alpha_n}{F_\nu}^{\alpha_2...\alpha_n}-\frac{n-1}{D-2}g_{\mu\nu}F_n^2)\cr &&\cr
\frac{1}{\sqrt{-g}}\partial_\mu(\sqrt{-g}e^{-2\phi}g^{\mu\nu}\partial_\nu\phi)&=&-\frac14\sum_n\frac{n-D/2}{2n!(1-D/2)}F_n^2\cr &&\cr
\partial_\mu(\frac{\sqrt{-g}}{n!}F^{\mu\alpha_2...\alpha_n})&=&0\,.\label{IIBeom}
\eea

One can construct this system by considering the $N_1$ number of D1-branes wrapped along the compact direction $y$ (a circle with radius $R$), the $N_5$ number of D5-branes wrapped along the $y$ direction and a four-torus $T^4$ where its coordinates are labeled by $z_i$ and its volume is $V_4$. In addition we have $p=N_p/R$ units of KK momentum along the $y$ direction in order to have a black hole with finite size area of the horizon. Then the solution to the equations of motion in (\ref{IIBeom}) is
\bea
ds^2&=&(f_1f_5)^{-\frac12}\bigg(-dt^2+dy^2+k(dt-dy)^2\bigg)\cr &&\cr
&+&(f_1f_5)^{\frac12}\bigg(dr^2+r^2d\Omega_3^2\bigg)+(\frac{f_1}{f_5})^{\frac12}\bigg(dz_i^2\bigg)\,,\cr &&\cr
e^{-2\phi}&=&\frac{f_5}{f_1}\,,\,\,\,\,
C^{(2)}_{ty}=(\frac{1}{f_1}-1)\,,\,\,\,\,C^{(6)}_{tyz_1...z_4}=(\frac{1}{f_5}-1)\,,\label{sol}
\eea
with three harmonic functions
\be
f_1=1+\frac{Q_1}{r^2}\,,\,\,\,\,f_5=1+\frac{Q_5}{r^2}\,,\,\,\,\,k=\frac{Q_p}{r^2}\,,\label{fk}
\ee
where $Q_1$, $Q_5$ and $Q_p$ are constants of integration an are related to the number of D-branes and KK momentum via
\bea
N_1&=&\frac{1}{16\pi G_{10}T_1}\int_{S^3\times T^4}*(F^{(3)}_{tyr})=\frac{V_{T^4}Q_1}{16\pi^4\a^3g_s}\,,\cr &&\cr
N_5&=&\frac{1}{16\pi G_{10}T_5}\int_{S^3}F^{(3)}_{\theta\phi\psi}=\frac{Q_5}{g_s\a}\,,\cr &&\cr
N_p&=&pR=\frac{V_{T^4}R^2Q_p}{16\pi^4g_s^2\a^4}\,,\label{n1n5}
\eea
and $p$ is the total ADM momentum in the $y$ direction and can be computed as the Noether charge of the gauge field corresponding to the off diagonal component $g_{ty}$.\footnote{Here we have considered $16\pi G_{10}=(2\pi)^7\alpha'^4 g_s^2$ and $T_p=2\pi/(2\pi\alpha'^{\frac12})^{p+1}g_s$.}

It is possible to assume that D5-branes are sources of magnetic fields. Therefore, it is enough to just consider a three form field strength with non vanishing components
\be
F^{(3)}_{tyr}=\frac{2r}{Q_1}\,,\,\,\,F^{(3)}_{\theta\phi\psi}=2Q_5\sin^2\theta\sin\phi\,.\label{f3}
\ee
In what follows we use this formalism for computing the entropy of $D1D5p$ black holes.
\section{Entropy function of D1D5p black hole at d=5}
The entropy function formalism is a useful tool for computing the entropy of black holes with near horizon $AdS_2\times S^{d-2}$ \cite{Sen:2005wa}. Using this fact which the near horizon geometry of $D1D5p$ black holes in $d=5$ is $AdS_2\times S^3$ we use this formalism for computing the entropy of $D1D5p$ black holes. 

We start with IIB supergravity in ten dimensions and compactify it on $S^1\times T^4$. This action in string frame is (see e.g. \cite{Kiem:1997qj}) 
\bea
S_5&=&\frac{1}{2\kappa_5^2}\int d^5x\sqrt{-g^{(5)}}e^{2\psi+\frac{\psi_1}{2}}\bigg(e^{-2\phi}R^{(5)}-\frac{1}{12} H_3^2-\frac14e^{-\psi_1}F_2^2-\frac14e^{-2\phi+\psi_1}C_2^2\cr &&\cr
&+&e^{-2\phi}[2\partial_\mu\psi\partial^\mu\psi_1-2\partial_\mu\phi\partial^\mu\psi_1-8\partial_\mu\phi\partial^\mu\psi+4\partial_\mu\phi\partial^\mu\phi+3\partial_\mu\psi\partial^\mu\psi]\bigg)\,,\label{act5}
\eea
where $2\kappa_5^2=2\pi RV_4/16\pi G_{10}$ is the five dimensional effective coupling. In the above relation ${\psi}$ and ${\psi_1}$ are single moduli for $T^4$ and $S^1$. $H_3$ is a three form magnetic field strength. $F_2$ is a two form electric field strength coming from compactifying the ten dimensional 3-form field strength and $C_2$ is a two form field strength for the gauge field corresponding to KK momentum around the $y$ direction.

The solution for the metric from equations of motion for this action is (see e.g. \cite{Peet:2000hn})

\be
ds_5^2=-\frac{(f_1f_5)^{-\frac12}}{1+k} dt^2+(f_1f_5)^\frac12\bigg(dr^2+r^2d\Omega_3^2\bigg)\,,\label{metr5}
\ee
where again we have the previous harmonic functions
\be
f_1=1+Q_1/r^2\,,\,\,\,\,f_5=1+Q_5/r^2\,,\,\,\,\,k=Q_p/r^2\,,
\ee
and also the same relations for constants of integration and number of $D$-branes
\be
N_1=\frac{V_{T^4}Q_1}{16\pi^4\a^3g_s}\,,\,\,\,\,N_5=\frac{Q_5}{g_s\a}\,,\,\,\,\,N_p=\frac{{V_{T^4}R^2Q_p}}{{16\pi^4g_s^2\a^4}}\,.\label{n1n5p}
\ee

The near horizon limit of the background (\ref{metr5}) is $AdS_2\times S^3$, so there must be an entropy function where its extremum value gives the entropy of the black hole. To see this we start from (\ref{metr5}) and deform its near horizon geometry to
\be
ds_5^2=v_1\bigg(-\frac{1}{Q_p\sqrt{Q_1Q_5}}r^4dt^2+\sqrt{Q_1Q_5}\frac{dr^2}{r^2}\bigg)+v_2\bigg(\sqrt{Q_1Q_5}d\Omega_3^2\bigg)\,,\label{nearh5}
\ee
with the following properties for the Riemann tensor components
\bea
&&R_{abcd}=\frac{4}{v_1\sqrt{Q_1Q_5}}(g_{ac}g_{bd}-g_{ad}g_{bc})\,,\,\,\,\,a,b,c,d=r,t,\cr &&\cr
&&R_{ijkl}=-\frac{1}{v_2\sqrt{Q_1Q_5}}(g_{ik}g_{jl}-g_{il}g_{jk})\,,\,\,\,\,i,j,k,l=\theta,\phi,\psi\,.
\eea
Because the horizon is a three sphere sitting at $r=0$, the form of metric indicates the following result for the Wald formula  
\be
S_{BH}=-8\pi\int dx_H\sqrt{g^H}\frac{\partial L}{\partial R_{trtr}}g_{tt}g_{rr}\,.
\ee
Now we introduce $f$ as integral of density Lagrangian over the horizon
\be
f=\frac{1}{2\kappa_5^2}\int dx_H \sqrt{-g}L\,,
\ee
and rescale $R_{trtr}$ component of the Riemann tensor in the Lagrangian density ($f$) by a factor of $\lambda$. Then the entropy can be written as
\be
S_{BH}=-\frac{\pi}{2r}\sqrt{Q_1Q_5Q_p}\frac{\partial{f}}{\partial \lambda}\bigg|_{\lambda=1}\,.
\ee
It is not hard to show that (see more details in the next section)
\be
\frac{\partial f}{\partial\lambda}\bigg|_{\lambda=1}=4\int dx_H \sqrt{-g}R_{trtr}\frac{\partial L}{\partial R_{trtr}}=f-e_i\frac{\partial{f}}{\partial e_i}\,.
\ee 
Finally we find the relation between entropy $S$ and the entropy function $F$ or Legendre transform of $f$ with respect to the electric field strengths $e_1$ and $e_2$ 
\be
S_{BH}=-\frac{\pi}{2r}\sqrt{Q_1Q_5Q_p}(f-e_i\frac{\partial{f}}{\partial e_i})=\frac{\pi}{2r}\sqrt{Q_1Q_5Q_p}F\,,
\ee
It remains to compute $F$. For doing this, in addition the following values for form fields we assume that all the scalars in the theory give a constant value near the horizon i.e. 
\bea
&&e^{-2\phi}=u_s\,,\,\,\,\,e^{\frac{\psi_1}{2}}=u_1\,,\,\,\,\,e^{2\psi}=u_2\,,\cr &&\cr
&&H_{\theta\phi\psi}=p\sin^2\theta\sin\phi\,,\,\,\,\,F_{tr}=e_1\,,\,\,\,\,C_{tr}=e_2\,.
\eea
Then the entropy function $F$ will be
\bea
F&=&\frac{4\pi^3RV_4}{16\pi G_{10}}r(Q_1Q_5)^\frac34 Q_p^{-\frac12}v_1v_2^\frac32 u_2\bigg\{2\frac{u_1u_s}{(Q_1Q_5)^\frac12}\frac{4v_2-3v_1}{v_1v_2}+\frac12\frac{u_1p^2}{(Q_1Q_5)^\frac32}\frac{1}{v_2^3}\cr &&\cr
&+&\frac12(\frac{e_1^2}{u_1}+e_2^2u_1^3u_s)\frac{Q_p}{v_1^2r^2}\bigg\}\,.
\eea
The equations of motion for metric and scalars now is translating to extremizing of the entropy function with respect to scalars $u_k$ and $v_i$ 
\be
\frac{\partial F}{\partial u_k}=0\,,\,\,\,\,k=1,2,s\,,\,\,\,\,\frac{\partial F}{\partial v_i}=0\,,\,\,\,\,i=1,2\,.
\ee
In addition to these equations there are equations of motion and Bianchi identities for the gauge fields which translate to $q_i=\partial f/\partial e_i$ where $q_i$ stands for the electric charges. This last type of equations of motion can be viewed from another point. In order to have the relations in (\ref{n1n5p}) we need the following results
\be
p=2Q_5\,,\,\,\,\,e_1=2r\frac{u_1}{u_2}\frac{v_1}{v_2^\frac32}(\frac{Q_1}{Q_p^2Q_5^3})^\frac14\,,\,\,\,\,e_2=2r\frac{1}{u_1^3u_2u_s}\frac{v_1}{v_2^\frac32}\frac{Q_p^\frac12}{(Q_1Q_5)^\frac34}\,.
\ee  
Finally the solution to the equations of motion gives
\bea
&&v_1=v_2=v\,,\,\,\,\,u_s=\frac{Q_5}{Q_1}\frac{1}{v^2}\,,\,\,\,\,u_1^2=\frac{Q_p}{\sqrt{Q_1Q_5}}v\,,\,\,\,u_2=\frac{Q_1}{Q_5}\,,\,\,\,\,\cr &&\cr  &&e_1=2r\frac{1}{Q_1}\,,\,\,\,\,e_2=2r\frac{1}{Q_p}\,,
\eea 
We do not need here to know the value of $v$ because by putting back all these values to the entropy function, it will be independent of the value $v$. Any way in the metric it is just an overall factor and can be set equal to one. We finally find the famous result for entropy of five dimensional $D1D5p$ black holes 
\be
S_{BH}=2\pi\sqrt{N_1N_5N_p}\,.
\ee
This result has been found with different methods either by direct computing the area of the horizon or by counting the number of degeneracy of the microstates.

\section{Entropy function of D1D5p at d=10}
In this section we want to find the entropy function of $D1D5p$ black holes in $d=10$. Although as we saw in the previous section that this computation is simple in $d=5$ because of simple near horizon geometry $AdS_2\times S^3$ but here we will show that for more general near horizon metrics other than $AdS_2\times S^{d-2}$ that the method of Sen works. We will show this in the leading and next leading order of $\a$ low energy effective Lagrangian for IIB string theory, but it seems that this is a general properties of all corrected Lagrangians coming from string theory. 
\subsection{Entropy function at leading order}
Now we start from ten dimensional metric and modify the near horizon geometry of $D1D5p$ system to the following form
\bea
ds^2&=&v_1\bigg(\frac{Q_p-r^2}{\sqrt{Q_1Q_5}}dt^2+\sqrt{Q_1Q_5}\frac{dr^2}{r^2}+\frac{Q_p+r^2}{\sqrt{Q_1Q_5}}dy^2-2\frac{Q_p}{\sqrt{Q_1Q_5}} dtdy\bigg)
\cr &&\cr
&+&v_2\bigg(\sqrt{Q_1Q_5}d\Omega_3^2+\sqrt{\frac{Q_1}{Q_5}}dz_i^2\bigg)\,.\label{metr}
\eea
One can see from this metric that in the case where $Q_p=0$ the geometry is $AdS_3\times S^3\times T^4$, for the above case the horizon is $S^1\times S^3\times T^4$ and sits at $r=0$. Another important property of this deformed metric is the following relations for the Riemann tensor components
\bea
R_{abcd}&=&\frac{1}{v_1\sqrt{Q_1Q_5}}(g_{ac}g_{bd}-g_{ad}g_{bc})\,,\,\,\,\,a,b,c,d=t,y,r\,,\cr &&\cr
R_{ijkl}&=&-\frac{1}{v_2\sqrt{Q_1Q_5}}(g_{ik}g_{jl}-g_{il}g_{jk})\,,\,\,\,\,i,j,k,l=\theta,\phi,\psi\,.\label{rgggg}
\eea
Now we assume that dilaton keeps a constant value. Also consider $D1$-branes as electric field sources and $D5$-branes as magnetic field sources with the following values for their field strength components 
\bea
e^{-2\phi}=u_s\,,\,\,\,\,F^{(3)}_{tyr}=e_1\,,\,\,\,\,F^{(3)}_{\theta\phi\psi}=p_1\,.\label{phif}
\eea
Because the geometry has been deformed we need again to compute the relation between number of D-branes, $N_i$ with constants of integrations, $Q_i$ and also the ADM momentum in the y direction. But we are not changing our configuration of D-branes and KK momentum so we keep equation (\ref{n1n5}) and just change the values of $u_s$, $e_1$ and $p_1$ as
\bea
u_s=\frac{1}{v_1^\frac12 v_2^\frac72}\frac{Q_5}{Q_1}\,,\,\,\,\,e_1=\frac{2r}{Q_1}\frac{v_1^\frac32}{v_2^\frac72}\,,\,\,\,\,p_1=2Q_5\sin^2\theta\sin\phi\,.\label{uep}
\eea
This is similar to what we have done in previous section as an alternative for equations of motion and Bianchi identities for gauge fields.

In order to compute the entropy function we need to know the Wald formula for this new background. The metric in (\ref{metr}) has an off diagonal component that makes the Wald formula much more complicated to compute. The general form of Wald formula for entropy is \cite{Visser:1993qa}
\be
S_{BH}=-4\pi\int_H dx_H \sqrt{g^H}\frac{\partial L}{\partial R_{\mu\nu\lambda\rho}}g_{\mu\lambda}^{\bot}g_{\nu\rho}^{\bot}\,,\label{wald}
\ee
where $g_{\mu\nu}^{\bot}$ denotes the projection onto subspace orthogonal to the horizon.
For our case the nonzero components of the orthogonal metric will be (see appendix A)
\bea
&&g_{tt}^{\bot}=-\frac{(r^2-Q_p)v_1}{\sqrt{Q_1Q_5}}\,,\,\,\,\,g_{rr}^{\bot}=\frac{\sqrt{Q_1Q_5}v_1}{r^2}\,,\cr &&\cr
&&g_{yy}^{\bot}=-\frac{Q_p^2v_1}{\sqrt{Q_1Q_5}(r^2-Q_p)}\,,\,\,\,\,g_{ty}^{\bot}=-\frac{Q_pv_1}{\sqrt{Q_1Q_5}}\,,\label{ortmetr}
\eea
and the Riemann tensor components that enter in the computation of the entropy are
\bea
&& R_{trtr}=-\frac{v_1(r^2-Q_p)}{\sqrt{Q_1Q_5}r^2}\,,\,\,\,\,R_{tyty}=-\frac{v_1r^4}{(Q_1Q_5)^{3/2}}\,,\cr &&\cr
&& R_{yryr}=\frac{v_1(r^2+Q_p)}{\sqrt{Q_1Q_5}r^2}\,,\,\,\,\,R_{tryr}=-\frac{Q_pv_1}{\sqrt{Q_1Q_5}r^2}\,,\label{riman}
\eea
knowing the above relations we can replace the orthonormal metric with Riemann tensors in the Wald formula (\ref{wald}) and rewrite the entropy as
\bea
S&=&\sum_{i=1}^4S_i\,,\cr &&\cr
S_1&=&-8\pi v_1 \sqrt{Q_1Q_5}\int_H dx_H  \sqrt{g^H}\frac{\partial L}{\partial R_{rtrt}}R_{rtrt}\,,\cr &&\cr
S_2&=&8\pi v_1 \sqrt{Q_1Q_5}\frac{Q_p^2}{r^4-Q_p^2}\int_H dx_H \sqrt{g^H}\frac{\partial L}{\partial R_{ryry}}R_{ryry}\,,\cr &&\cr
S_3&=&-16\pi v_1 \sqrt{Q_1Q_5}\int_H dx_H  \sqrt{g^H}\frac{\partial L}{\partial R_{tryr}}R_{tryr}\,,\cr &&\cr
S_4&=&-8\pi \int_H dx_H \sqrt{g^H}\frac{\partial L}{\partial R_{tyty}}(g_{tt}^{\bot}g_{yy}^{\bot}-(g_{ty}^\bot)^2)=0\,.\label{ent}
\eea
We again introduce function $f$ as integral of Lagrangian density over the horizon
\be
f=\int_H dx_H \sqrt{-g}L\,,\label{f}
\ee
and rescale the Riemann tensor components in (\ref{riman}). Notice that although the $S_4$ is zero but for the future considerations we need to rescale the $R_{tyty}$ component as well
\be
R_{rtrt}\rightarrow\lambda_1R_{rtrt}\,,\,\,\,\,R_{ryry}\rightarrow\lambda_2R_{ryry}\,,\,\,\,\,R_{tryr}\rightarrow\lambda_3R_{tryr}\,,\,\,\,R_{tyty}\rightarrow\lambda_4R_{tyty}\,.\label{rescal}
\ee 
The rescaled Lagrangian behaves as ${\partial L_{\lambda}}/{\partial\lambda_i}=R^{(i)}_{\mu\nu\lambda\rho}{\partial L_\lambda}/{\partial R^{(i)}_{\mu\nu\lambda\rho}}$ with no summation on the right hand side for $i$ ($i=1,2,3,4$). 

Using the relation $\sqrt{g^H}=\sqrt{-g}v_1^{-1}{(1+Q_p/r^2)^{1/2}}$ we find the following equation for the rescaled function $f_\lambda$
\be
\frac{\partial f_{\lambda_i}}{\partial\lambda_i}\bigg|_{\lambda_i=1}=v_1(1+\frac{Q_p}{r^2})^{-1/2}\int_H dx_H \sqrt{g^H}R_{\mu\nu\lambda\rho}^{(i)}\frac{\partial L}{\partial R_{\mu\nu\lambda\rho}^{(i)}}\,,\label{df}
\ee
replacing this relation into the entropy formula (\ref{ent}) we find
\be
S=-2\pi \sqrt{Q_1Q_5(1+\frac{Q_p}{r^2})}\bigg\{\frac{\partial f_{\lambda_1}}{\partial\lambda_1}-\frac{Q_p^2}{r^4-Q_p^2}\frac{\partial f_{\lambda_2}}{\partial\lambda_2}+\frac{\partial f_{\lambda_3}}{\partial\lambda_3}\bigg\}_{\lambda_1=\lambda_2=\lambda_3=1}\,.\label{entf}
\ee
Looking to the equation (\ref{riman}) shows that the Riemann tensor components are divided in to two parts. All the components in the first part are proportional to $v_1$. In a general Lagrangian these Riemann tensors are contracting with those components of the inverse metric which have $r$, $t$ and $y$ indices, where all are proportional to $v_1^{-1}$, so in general we have the following behavior  
\bea
&&\lambda_1 R_{rtrt}g^{rr}g^{tt}\sim\lambda_1 v_1^{-1}\,,\,\,\,\,
\lambda_2 R_{ryry}g^{rr}g^{yy}\sim\lambda_2 v_1^{-1}\,,\,\,\,\cr &&\cr
&&\lambda_3 R_{tryr}g^{rr}g^{ty}\sim\lambda_3 v_1^{-1}\,,\,\,\,\,
\lambda_4 R_{tyty}g^{tt}g^{yy}\sim\lambda_4 v_1^{-1}\,,\label{sym1}
\eea
which means that everywhere in the Lagrangian for every $\lambda_i$ there exist a $v_1^{-1}$. In addition the electric field strength enters as $\sqrt{g^{rr}(g^{tt}g^{yy}-(g^{ty})^2)}F_{tyr}\sim e_1 v_1^{-1}$ but the magnetic field contribution has not any dependence on $v_1$. Also one can check that all covariant derivatives of Riemann tensor and field strength are vanishing. These facts help us to guess the behavior of $f_\lambda$ as a function of scalars, electric and magnetic field strengths  
\be
f_\lambda(u_s,v_1,v_2,e_1,p_1)=v_1^\frac32 g(u_s,v_2,\lambda_i v_1^{-1},e_1 v_1^{-1},p_1)\,,\label{fg}
\ee
where $g$ is a general function. Simply one can show that
\be
\sum_{i=1}^4\lambda_i\frac{\partial f_\lambda}{\partial\lambda_i}=\frac32(f_\lambda-e_1\frac{\partial f_\lambda}{\partial e_1})-v_1\frac{\partial f_\lambda}{\partial v_1}\,.\label{partials}
\ee
As we see although in (\ref{entf}) the derivative with respect to $\lambda_4$ has not appeared but by using the equations of motion we can write the entropy in (\ref{entf}) as
\be
S=-2\pi \sqrt{Q_1Q_5(1+\frac{Q_p}{r^2})}\bigg\{\frac32(f-e_1\frac{\partial f}{\partial e_1})-(\frac{\partial f}{\partial\lambda_4}+\frac{r^4}{r^4-Q_p^2}\frac{\partial f}{\partial\lambda_2})\bigg\}\,.\label{entf2}
\ee
which brings the $\lambda_4$ into our game. In order to get ride of this complicated formula we need some simplifying relations.
It is possible to show that we have the following relations 
\bea
&&\frac{\partial f}{\partial \lambda_1}=\frac{\partial f}{\partial \lambda_2}\,,\,\,\,\,\cr &&\cr
&&2\frac{\partial f}{\partial \lambda_1}+\frac{\partial f}{\partial \lambda_3}=2\frac{\partial f}{\partial \lambda_4}\cr &&\cr
&&\frac{\partial f}{\partial \lambda_3}/\frac{\partial f}{\partial \lambda_2}=\frac{2Q_p^2}{r^4-Q_p^2}\,,\label{simps}
\eea
we have found this relation for supergravity action but they are also correct for the next leading order of corrections for the Lagrangian. It seems that these relations are general properties of Lagrangians in IIB theory (with symmetric background metric). For example the first one can be understood from these fact that there is a symmetry between indices $t$ and $y$ every where in the Lagrangian.
Using above equations one can verify a very simple formula for entropy in terms of Legendre transform of function $f$ with respect to electric field strengths 
\be
S=\pi\sqrt{Q_1Q_5(1+\frac{Q_p}{r^2})}(e_1\frac{\partial f}{\partial e_1}-f)\label{leg}
\ee
We can introduce entropy function $F=e_1\frac{\partial f}{\partial e_1}-f$ and extremize it with respect to scalar moduli
\be
\frac{\partial F}{\partial u_s}=0\,,\,\,\,\,\frac{\partial F}{\partial v_i}=0\,,\,\,\,\,i=1,2\,.\label{Feom}
\ee 
Inserting the values of metric, dilaton and the field strength into the entropy function and integrating over the horizon $S^1\times S^3\times T^4$ gives
\be
F=\frac{4\pi^3RV_4}{16\pi G_{10}}rQ_1^{\frac32}Q_5^{-\frac12}v_1^{\frac32}v_2^{\frac72}\bigg[\frac{6u_s (v_2-v_1)}{(Q_1Q_5)^\frac12 v_1v_2}+\frac{Q_5^{\frac12}}{Q_1^{\frac32}}(\frac{2}{v_2^3}+\frac{e_1^2Q_1^2}{2r^2v_1^3})\bigg]\,.\label{f1}
\ee 
If one replace the values in (\ref{uep}) in to the entropy function and extremize it in terms of $v_1$ and $v_2$, then the values of scalar moduli will be
\be
v_1=v_2=1\,,\,\,\,\,u_s=\frac{Q_5}{Q_1}\,,
\ee
inserting this values into $F$ and using (\ref{n1n5}) we finally find the value for entropy as before 
\be
S=2\pi\sqrt{N_1N_5N_p}\,,\label{entropy}
\ee 
\subsection{Entropy function at next leading order}
In this subsection we try to compute the corrections to entropy function due to $\alpha'$ corrections coming from string theory three level scattering amplitude computations. We just proposed corrections to the gravity part of the Lagrangian but as we mentioned in previous section all the covariant derivatives of Riemann tensor and the field strengths are vanishing so it seems that there is not any other contribution to our results. These corrections are \cite{Grisaru:1986vi}
\bea
L_{corr}&=&\gamma e^{-2\phi}(L_1-2L_2+\lambda L_3)\,,\cr &&\cr
L_1&=&R^{hmnk}R_{pmnq}{R_h}^{rsp}{R^q}_{rsk}+\frac12 R^{hkmn}R_{pqmn}{R_h}^{rsp}{R^q}_{rsk}\,, \cr &&\cr
L_2&=&R^{hk}\bigg(\frac12R_{hnpk}R^{msqn}{R_{msq}}^p+\frac14R_{hpmn}{R_k}^{pqs}{R_{qs}}^{mn}+R_{hmnp}{R_{kqs}}^pR^{nqsm}\bigg)\,,\cr &&\cr
L_3&=&R\bigg(\frac14R_{hpmn}R^{hpqs}{R_{qs}}^{mn}+R_{hmnp}{{R^h}_{qs}}^{p}R^{nqsm}\bigg)\,,\label{L123}
\eea
where $\gamma=\frac18\zeta(3)(\a)^3$. Here the coefficient $\lambda$ is arbitrary due to the ambiguity in the field redefinition of the metric. Additionally it is possible to go to another coordinate system with vanishing Ricci tensor so that the Lagrangian $L_2$ vanishes and in the Lagrangian $L_1$ every Riemann tensors are replaced by Weyl tensors. But for the moment we consider all these three terms in our calculations.

The discussions of the previous subsection are valid here and all the steps is satisfying for next leading order of computations. The only change is correction to the entropy function. The corrected entropy function is
\bea
F&=&\frac{4\pi^3RV_4}{16\pi G_{10}}rQ_1^{\frac32}Q_5^{-\frac12}v_1^{\frac32}v_2^{\frac72}\bigg[\frac{6u_s (v_2-v_1)}{(Q_1Q_5)^\frac12 v_1v_2}+\frac{Q_5^{\frac12}}{Q_1^{\frac32}}(\frac{2}{v_2^3}+\frac{e_1^2Q_1^2}{2r^2v_1^3})\cr &&\cr
&+&6\gamma u_s\frac{v_1^4+v_2^4-6\lambda Q_5^\frac12(v_1^4-v_2v_1^3+v_2^4-v_1v_2^3)}{v_1^4v_2^4Q_1^2Q_5^2}\bigg]\,,\label{efc}
\eea
and the solutions to the equations of motion from extremizing the entropy function are
\bea
&&v_1=1+\frac{25}{4}\frac{\gamma}{(Q_1Q_5)^\frac32}\,,\,\,\,\,
v_2=1+\frac{9}{4}\frac{\gamma}{(Q_1Q_5)^\frac32}\,,\cr &&\cr
&&u_s=\frac{Q_5}{Q_1}(1-11\frac{\gamma}{(Q_1Q_5)^\frac32})\,,\,\,\,\,
e_1=\frac{2r}{Q_1}(1+\frac{3}{2}\frac{\gamma}{(Q_1Q_5)^\frac32})\,,\label{solc}
\eea
where as one see, obviously there is not any dependence on the parameter $\lambda$ which means that the entropy is independent of ambiguities. The corrected form of entropy now becomes
\be
S=2\pi\sqrt{N_1N_5N_p}\bigg(1+3\gamma(\frac{(2\pi)^3V_{4}}{16\pi G_{10}N_1N_5})^\frac32\bigg)\,,\label{ec}
\ee
one can see from this answer that the corrected entropy is symmetric under changing $N_1$ and $N_5$ and also it is related to other parameters, the volume of four-torus and ten dimensional effective coupling.
\section{Discussion}
In section 3 we see that by starting from the reduced action $IIB/S^1\times T^4$ we can find the entropy function for 3-charged $D1D5p$ black holes with steps similar to \cite{Sen:2005wa}. In our case, the orthogonal geometry to the horizon is $AdS_2$ space and horizon is a three sphere. In five dimensional reduced action, gravity is coupled to a three form magnetic field strength related to the presence of $D5$-branes. It is also coupled to two electric field strengths, one is originated from the existence of $D1$-branes and the other is a gauge field corresponding to KK momentum. The entropy function is the Legendre transform of function $f$, the integral of density Lagrangian over the horizon, with respect to these two electric field strengths. The difference here is the coefficient of proportionality between entropy and entropy function which in the case \cite{Sen:2005wa} is $2\pi$ and in our case is 
\be
S=\frac{\pi}{2r}\sqrt{Q_1Q_5Q_p}F 
\ee
which obviously depends on the form metric which we start with. By extremizing the entropy function we find that all the scalars in the theory gives a constant value. Putting back these values to the entropy function gives the well known entropy relation for $D1D5p$ black holes
\be
S_{D1D5p}=2\pi\sqrt{N_1N_5N_p}\,.
\ee
These results has been found with different methods either by direct computing the area of the horizon or by counting the number of degeneracy of the microstates.
In section four, we look to the problem from ten dimensional point of view. We consider the near horizon geometry of $D1D5p$ black holes. This geometry has two parts, first one, $M_3$ is a boosted $AdS_3$ space and second one is $S^3\times T^4$. We deform this geometry by choosing arbitrary constant values for radii of both spaces. These are two scalars of the theory that must be found from extremizing the entropy function. The other scalar of the theory is dilaton. In addition we have an electric and magnetic fields due to the presence of $D1$ and $D5$-branes as before. 
The form of $M_3$ makes a complicated Wald formula which depends on the projected metric onto subspace orthogonal to the horizon ($S^1\times S^3\times T^4$). Using the previous method for finding the entropy function and some useful relations (\ref{simps}) we find again the relation between entropy and entropy function in ten dimensions
\be
S=\pi\sqrt{Q_1Q_5(1+\frac{Q_p}{r^2})}F.
\ee
One can show that for the next leading order of corrections to IIB effective action which are $\a^3R^4$ type corrections this relation is very useful and gives very easily correction to the entropy. The point here is that the equation (\ref{simps}) is valid also in this level of computations. Therefore this suggests that it could be a general property of all terms in the action of IIB (with a symmetric background metric). 

The extremization of entropy function with respect to scalars gives a set of complicated equations which can be solve perturbatively in terms of $\a^3$. The correction to entropy of $D1D5p$ then will be
\be
S=2\pi\sqrt{N_1N_5N_p}\bigg(1+3\gamma(\frac{(2\pi)^3V_{4}}{16\pi G_{10}N_1N_5})^\frac32\bigg)\,.
\ee
It is also worth to find this correction from the five dimensional point of view by reducing the corrections to effective Lagrangian of IIB on $S^1\times T^4$. If we just look at gravitational terms from these reduction we can see the same behavior (functionality in terms of $N_1$ and $N_5$). The correct coefficient can be found by considering all the reduced parts of the Lagrangian.

\section*{Acknowledgments}
I would like to thank Mohsen Alishahiha for his useful comments and contribution at the early stages of this paper. I would like to thank Ashok Sen and Da-Wei Pang for their valuable comments. Most of this work has been done in CPHT at Ecole Polytechnique, I would like also to thanks Farhad Ardalan and Elias Kiritsis for giving me such an opportunity. This work was partially supported by ANR grant NT05-1-41861,
INTAS grant, 03-51-6346, RTN contracts MRTN-CT-2004-005104 and MRTN-CT-2004-503369,
CNRS PICS 2530 and 3059 and by a European Excellence Grant, MEXT-CT-2003-509661.

\appendix
\section{Orthogonal metric to horizon}

In order to find the orthogonal metric to horizon we need to know the unit normal and tangent vectors to the horizon. Choosing 
\be
N_{t}=(A,0,B,\vec{0})\,,\,\,\,\,N_{r}=(0,E,0,\vec{0})\,,\,\,\,\,T_{y}=(C,0,D,\vec{0})\,,
\ee
where we have considered the coordinates as $t,r,y$ and vector is denoted for coordinates in $S^3\times T^4$. For the general form of the metric
\be
ds^2=g_{tt}dt^2+g_{yy}dy^2+g_{ty}dtdy+g_{rr}dr^2+d\vec{x}^2\,,
\ee
the normal and tangent vectors to horizon at $S^1\times S^3\times T^4$ will be
\bea
N_{t}&=&\sqrt{\frac{g^{yy}}{g^{tt}g^{yy}-(g^{ty})^2}}(1,0,-\frac{g^{yt}}{g^{yy}},\vec{0})\,,\,\,\,\,
N_{r}=(0,\frac{1}{\sqrt{g^{rr}}},0,\vec{0})\,,\cr &&\cr
T_{y}&=&(0,0,\frac{1}{\sqrt{g^{yy}}},\vec{0})\,,
\eea
using the definition of orthogonal metric
\be
g_{\mu\nu}^{\bot}=(N_t)_\mu(N_t)_\nu+(N_r)_\mu(N_r)_\nu\,,
\ee 
we find the following values for the orthonormal metric to the horizon of $D1D5p$ black hole as
\bea
&&g_{tt}^{\bot}=-\frac{(r^2-Q_p)v_1}{\sqrt{Q_1Q_5}}\,,\,\,\,\,g_{rr}^{\bot}=\frac{\sqrt{Q_1Q_5}v_1}{r^2}\,,\cr &&\cr
&&g_{yy}^{\bot}=-\frac{Q_p^2v_1}{\sqrt{Q_1Q_5}(r^2-Q_p)}\,,\,\,\,\,g_{ty}^{\bot}=-\frac{Q_pv_1}{\sqrt{Q_1Q_5}}\,.
\eea

\end{document}